\documentclass[times, twocolumn,authoryear]{aastex631}
\usepackage{CJK}
\usepackage{graphicx}
\usepackage{enumerate}
\usepackage{amssymb, amsmath}
\usepackage{color}
\usepackage{ulem}
\usepackage{appendix}
\usepackage{comment}
\usepackage{natbib}

\usepackage[nodayofweek]{datetime}
\newdateformat{monthyearday}{%
  \THEYEAR\ \monthname[\THEMONTH] \twodigit{\THEDAY}}
  
\submitjournal{AAS Journals}
\shorttitle{HIP~53005C}
\shortauthors{Uyama et al.}

\begin{document}

\title{Discovery of a Low-Mass Companion to the Accelerating Star HIP 53005 with Strongly Conflicting Mass Estimates}

\correspondingauthor{Taichi Uyama}
\email{taichi.uyama.astro@gmail.com}

\author[0000-0002-6879-3030]{Taichi Uyama}
    \affiliation{Astrobiology Center, 2-21-1 Osawa, Mitaka, Tokyo 181-8588, Japan}
    \affiliation{National Astronomical Observatory of Japan, 2-21-1 Osawa, Mitaka, Tokyo 181-8588, Japan}

\author[0000-0002-7405-3119]{Thayne Currie}
    \affiliation{Department of Physics and Astronomy, University of Texas-San Antonio, 1 UTSA Circle, San Antonio, TX}
    \affiliation{Subaru Telescope, National Astronomical Observatory of Japan, National Institutes of Natural Sciences, 650 North A'oh$\bar{o}$k$\bar{u}$ Place, Hilo, HI 96720, USA}

\author[0000-0002-6618-1137]{Jerry W. Xuan}
\altaffiliation{51 Pegasi b Fellow}
\affiliation{Department of Astronomy, California Institute of Technology, Pasadena, CA 91125, USA}
\affiliation{Department of Earth, Planetary, and Space Sciences, University of California, Los Angeles, CA 90095, USA}

\author[0000-0002-4918-0247]{Robert De Rosa}
    \affiliation{European Southern Observatory, Alonso de C$\acute{o}$rdova 3107, Vitacura, Santiago, Chile}

\author[0000-0002-4677-9182]{Masayuki Kuzuhara}
    \affiliation{Astrobiology Center, 2-21-1 Osawa, Mitaka, Tokyo 181-8588, Japan}
    \affiliation{National Astronomical Observatory of Japan, 2-21-1 Osawa, Mitaka, Tokyo 181-8588, Japan}

\author[0000-0001-8892-4045]{Minghan Chen}
    \affiliation{Department of Physics, University of California, Santa Barbara, Santa Barbara, CA 93106, USA}

\author[0000-0002-3122-6809]{Vito Squicciarini}
    \affiliation{LESIA, Observatoire de Paris, Université PSL, CNRS, Sorbonne Université, Université Paris Cité, 5 place Jules Janssen, 92195 Meudon, France}
    \affiliation{INAF – Osservatorio Astronomico di Padova; Vicolo dell’Osservatorio 5, I-35122 Padova, Italy}

\author[0000-0002-5627-5471]{Charles Beichman}
    \affiliation{NASA Exoplanet Science Institute, 1200 E. California Blvd., Pasadena, CA 91125, USA}
    \affiliation{Infrared Processing and Analysis Center, California Institute of Technology, 1200 E. California Blvd., Pasadena, CA 91125, USA}
\author[0000-0003-2630-8073]{Timothy D. Brandt}
    \affiliation{Space Telescope Science Institute, 3700 San Martin Drive, Baltimore, MD 21218, USA}
\author[0000-0003-4514-7906]{Vincent Deo}
    \affiliation{Subaru Telescope, National Astronomical Observatory of Japan, National Institutes of Natural Sciences, 650 North A'oh$\bar{o}$k$\bar{u}$ Place, Hilo, HI 96720, USA}
\author[0000-0002-1097-9908]{Olivier Guyon}
    \affiliation{Subaru Telescope, National Astronomical Observatory of Japan, National Institutes of Natural Sciences, 650 North A'oh$\bar{o}$k$\bar{u}$ Place, Hilo, HI 96720, USA}
    \affiliation{Steward Observatory, University of Arizona, Tucson, AZ 85721, USA}
    \affiliation{Astrobiology Center, 2-21-1 Osawa, Mitaka, Tokyo 181-8588, Japan}
\author[0000-0003-3618-7535]{Teruyuki Hirano}
    \affiliation{Astrobiology Center, 2-21-1 Osawa, Mitaka, Tokyo 181-8588, Japan}
\author[0000-0001-8345-593X]{Markus Janson}
    \affiliation{Department of Astronomy, Stockholm University, AlbaNova University Center, SE-10691, Stockholm, Sweden}
\author[0000-0003-2232-7664]{Michael C. Liu}
    \affiliation{Institute for Astronomy, University of Hawai'i, 2680 Woodlawn Drive, Honolulu HI 96822}
\author[0000-0002-8895-4735]{Dimitri Mawet}
    \affiliation{Department of Astronomy, California Institute of Technology, 1200 E. California Blvd., Pasadena, CA 91125, USA}
    \affiliation{Jet Propulsion Laboratory, California Institute of Technology, 4800 Oak Grove Dr., Pasadena, CA 91109, USA}
\author[0000-0002-3047-1845]{Julien Lozi}
    \affiliation{Subaru Telescope, National Astronomical Observatory of Japan, National Institutes of Natural Sciences, 650 North A'oh$\bar{o}$k$\bar{u}$ Place, Hilo, HI 96720, USA}
\author[0000-0003-4698-6285]{Stevanus Nugroho}
    \affiliation{Department of Earth and Planetary Sciences, School of Science, Institute of Science Tokyo, 2-12-1 Ookayama, Meguro, Tokyo 152-8551, Japan}
\author[0000-0002-6510-0681]{Motohide Tamura}
    \affiliation{Astrobiology Center, 2-21-1 Osawa, Mitaka, Tokyo 181-8588, Japan}
\author[0000-0003-4018-2569]{Sebastien Vievard}
    \affiliation{Subaru Telescope, National Astronomical Observatory of Japan, National Institutes of Natural Sciences, 650 North A`oh$\bar{o}$k$\bar{u}$ Place, Hilo, HI 96720, USA}

\author[0000-0001-7967-9922]{Danielle Bovie}
    \affiliation{Department of Physics and Astronomy, University of Texas-San Antonio, 1 UTSA Circle, San Antonio, TX}
\author[0000-0003-4676-0251]{Yasunori Hori}
    \affiliation{Graduate School of Environmental Life, Natural Science and Technology, Okayama University, 3-1-1 Tsushimanaka, Kita-ku, Okayama, Okayama 700-8530, Japan}
\author[0000-0003-3309-9134]{Hajime Kawahara}
    \affiliation{Department of Space Astronomy and Astrophysics, ISAS/JAXA, 3-1-1, Yoshinodai, Sagamihara, Kanagawa, 252-5210, Japan}
    \affiliation{Department of Astronomy, Graduate School of Science, The University of Tokyo, 7-3-1 Hongo, Bunkyo-ku, Tokyo 113-0033, Japan}
\author[0000-0001-6181-3142]{Takayuki Kotani}
    \affiliation{Astrobiology Center, 2-21-1 Osawa, Mitaka, Tokyo 181-8588, Japan}
    \affiliation{National Astronomical Observatory of Japan, 2-21-1 Osawa, Mitaka, Tokyo 181-8588, Japan}
    \affiliation{Department of Astronomical Science, The Graduate University for Advanced Studies, SOKENDAI, 2-21-1Osawa, Mitaka, Tokyo 181-8588, Japan}
\author[0000-0002-6845-9702]{Yiting Li}
    \affiliation{Department of Astronomy, University of Michigan, 1085 S. University, Ann Arbor, MI 48109, USA}
\author[0000-0003-0774-6502]{Jason Wang}
    \affiliation{Center for Interdisciplinary Exploration and Research in Astrophysics (CIERA) and Department of Physics and Astronomy, Northwestern University, Evanston, IL
60208, USA}


\begin{abstract}
We present the discovery of a low-mass companion located at $\rho$ $\sim$ 0\farcs{}85 ($r_{\rm proj} \approx 62~au$) from the early-type 1.2 Gyr-old star HIP 53005 using direct imaging data from the Subaru and Keck Telescopes and astrometry from the Hipparcos-Gaia Catalog of Accelerations.  The companion, HIP 53005 C, is a component of a multiple system also including a $\approx$ 12\farcs{}4-separation M dwarf companion inducing a negligible proper motion acceleration.   HIP~53005 C's position on color-magnitude diagrams, the fit of its spectral energy distribution to atmosphere models, and its location on an empirical mass-magnitude diagram all suggest that it lies at the M/L transition and near the hydrogen-burning limit ($\sim80~M_{\rm Jup}$).  However, our orbital fitting combining direct-imaging relative astrometry with proper motion acceleration favors a much higher dynamical mass of $\sim185\ M_{\rm Jup}$.  An additional unseen, more closely-orbiting companion below the detection limit (at $\rho\lesssim0\farcs2$)) may explain this discrepancy.  Alternatively, HIP~53005C could be a low-mass binary like Gliese~229Bab, making this system an intriguing laboratory for studying multiple star formation.
\end{abstract}


\section{Introduction} \label{sec: Introduction}

Direct imaging observations of brown dwarfs and jovian exoplanets offer fundamental insights into the formation, evolution, and atmospheric properties of substellar companions
\citep[e.g.][]{Currie2023-PP7}. Advancements in ground-based instrumentation, notably extreme adaptive optics systems (extreme AO), together with target selection based on indirect detection techniques (e.g., astrometry, radial velocity), have significantly enhanced the detection efficiency of such companions \citep[e.g., radial velocity trends, Hipparcos–Gaia proper motion accelerations;][]{Crepp2014,Brandt2020}. These approaches have yielded numerous new substellar detections \citep[e.g.][]{Currie2023,Currie2026,DeRosa2023,ElMorsy2025,Franson2023,Mesa2023}, representing a substantial improvement over earlier blind surveys \citep[e.g.][]{Nielsen2019,Currie2023-PP7}.

Combining direct imaging and precision astrometry further improves our ability to fully characterize substellar companions on wide orbits.   
Mass estimates based on direct imaging data alone rely on using a companion's luminosity to infer its mass given an estimated age.  However, this mapping from brightness to mass depends sensitively on precise stellar age estimates and relies on luminosity evolution models affected by uncertainties in opacities, cloud physics, and initial conditions \citep[e.g., hot-, warm-, and cold-start scenarios;][]{Spiegel2012}, especially at young ages and low surface gravities\footnote{Additionally, the typical temporal sampling of companion astrometry from direct imaging alone is very small compared to the decades to the objects' centuries-long orbital periods, precluding precisely-derived orbital parameters. }  However, jointly modeling the relative astrometry of companions from imaging data with the star's absolute astrometry from sources like Hipparcos and Gaia can yield direct dynamical masses \citep{Brandt2019a}.  

Independent dynamical mass measurements provide indispensable empirical benchmarks for calibrating theoretical models. 
Dynamical masses for companions with strong proper motions monitored by imaging observations over the course of several years can be measured to better than $\sim$10\% precision \citep[e.g.][]{Brandt2019a,GBrandt2021}.   These precise masses can provide key tests of luminosity evolution models \citep[e.g.][]{GBrandt2021}.  Systems with significant discrepancies between dynamical masses and luminosity evolution-estimated masses may hint at the presence of additional companions in the systems \citep[e.g.][]{Xuan2024}.
A key objective of ongoing direct imaging efforts is therefore to expand the sample of detected companions while constraining their dynamical masses and orbital parameters through synergistic analyses combining direct and indirect techniques.

In this paper, we present the discovery of a low-mass companion (HIP 53005 C) orbiting at $\sim$ 0\farcs{}85 from the early-type primary within the HIP 53005 multiple system using direct imaging data from the Subaru Coronagraphic Extreme Adaptive Optics Project \citep{Jovanovic2015} and Keck/NIRC2 coupled with astrometry from the \textit{Hipparcos Gaia Catalog of Accelerations} \citep{HGCA-GaiaEDR3}.   While the companion's photometry and spectrum suggests an object with a mass near the hydrogen-burning limit, its dynamical mass is $\approx$ a factor of two higher.  This discrepancy may hint at additional companions in the system or that HIP 53005 C is itself a binary, much like Gl 229B \citep{Xuan2024}.

\section{System Properties \label{sec: System Properties}}
HIP~53005 lies at a distance of $\approx73~pc$ \citep{GaiaDR3} and is categorized as a metallic-line or an Am star \citep[][]{Renson1991,Renson2009};\cite{McGahee2020} updated the classification as kA6/hF0/mF3(III)Sr (its $B-V$ color corresponds to $\sim1.7 M_\odot$, see Figure~5 in the reference).  
BANYAN $\Sigma$ \citep{gagne18} does not suggest a membership of HIP~53005 to any known young association or moving group, and previous studies did not provide an accurate age estimation. Therefore we estimated the age of this system with isochrone fitting (see Section~\ref{sec: System Age}). 

The Hipparcos-Gaia (EDR3) Catalog of Accelerations \citep{HGCA-GaiaEDR3} shows that HIP 53005 has a strong proper motion anomaly ($\chi^2$ $\sim$ 66.1), significant at the $7.84\sigma$ level. Its Renormalised Unit Weight Error (RUWE\footnote{An index of how a source is well fit with a single object.}) is $\sim$ 1 as 1.009, consistent with a single star solution, disfavoring the scenario that this anomaly comes from a massive, closely-orbiting stellar companion \citep{Gaia-EDR3}.  Thus, we targeted this system for follow-up AO imaging to detect the companion responsible for the star's acceleration.

\section{Data} \label{sec: Data}

\subsection{Subaru and Keck Observations} \label{sec: Subaru and Keck Observations}
\subsubsection{Observations}

We observed HIP 53005 with Subaru/SCExAO+CHARIS and Keck/NIRC2 in multiple epochs between 2022 and 2024 (see Table \ref{tab: observations}).  The SCExAO/CHARIS data were taken in the low-resolution integral field spectroscopic mode covering the $JHK$ bandpasses \citep[${\mathcal R}\sim16$, $16.10\pm0.04$~mas/pixel;][]{Groff2015,Currie2026}; Keck/NIRC2 consist of $K^\prime$ and $L^\prime$-band imaging \citep[$9.971\pm0.004$~mas/pixel;][]{Service2016}. 
All data were taken in angular differential imaging mode \citep[ADI;][]{Marois2006}, allowing the field of view to rotate on the detector plane with time.

To estimate a system age from the primary's rotation rate, we also obtained Subaru/IRD spectra of the central star in the $Y$ to $H$ bands (0.95~$\mu$m--1.75$\mu$m, $R\sim70000$) . 
Our data consist of 8 IRD sequences on 2022 March 24$^{\mathrm{th}}$ UT with exposure times of 900-1900 s resulting in S/N of 72-205 at 1~$\mu$m. For wavelength calibration, we used Th-Ar spectra taken on 2022 March 21$^{\mathrm{st}}$ UT.

\begin{deluxetable*}{cccccccc}
\label{tab: observations} 
\tablecaption{Observing Logs of High-contrast Imaging}
\tablehead{
\colhead{Instrument, Filter} & \colhead{Date [UT]} & DIMM seeing & \colhead{$t_{\rm int}$ [sec]}& \colhead{$N_{\rm exp}$} & \colhead{Field rotation} & \colhead{Reduction} & \colhead{SNR of C}
}
\startdata
        Subaru/CHARIS, $JHK$ & 2021 May 9 & $\sim0\farcs6-0\farcs8$ & 30.98 & 71 &  104\fdg3 & ADI & 198$^a$ \\
        Keck/NIRC2+pyWFS, $L^\prime$ & 2021 May 21 & \dots & 30 & 35 & 29\fdg8 & ADI & 30 \\ 
        Subaru/CHARIS, $JHK$ & 2022 February 21 & $\sim0\farcs6-0\farcs8$ & 30.98 & 26 &  61\fdg4 & ADI & 71$^a$ \\
        Keck/NIRC2+SHWFS$^b$, $L^\prime$ & 2022 April 20 & $\sim0\farcs3-0\farcs6$ & 30 & 20 & 63\fdg9 & ADI & 32 \\ 
        Subaru/CHARIS, $JHK$ & 2023 February 5 & \dots & 30.98 & 41 & 3\fdg8 & SDI & 25$^a$ \\
        Keck/NIRC2+pyWFS, $L'$ & 2024 January 22 & \dots & 17.86 & 30 & 65\fdg1 & ADI & 47 \\ 
        Keck/NIRC2+SHWFS, $K'$ & 2025 May 11 & $\sim0\farcs4-0\farcs8$ & 2.1 & 150$^c$ & 37\fdg6 & ADI & 45 
\enddata
\tablecomments{$^a$ The SNR is calculated from the wavelength-combined CHARIS image. 
$^b$ For the NIRC2 data, we did not use a coronagraph except for the 2022 April data, where we used a vortex coronagraph \citep{Serabyn2017}. $^c$ We obtained more exposures but some of the data were taken when the target was close to zenith, causing fast field rotation and an elongated PSF of the companion. We therefore dropped such exposures from the whole exposures.
}
\end{deluxetable*}

\subsubsection{Data Reduction: CHARIS and NIRC2}
Subaru/CHARIS is the integral field unit and data calibration requires more steps than NIRC2.

For CHARIS data, we first extracted the data cubes from the raw detector reads using the pipeline described in \citet{Brandt2017}.  For subsequent processing, we used CHARIS Data Processing Pipeline \citep[DPP;][]{Currie2020-DPP}, performing background subtraction and image registration based on the position of satellite spots generated by the SCExAO deformable mirror as a reference for photometry and astrometry of the coronagraphic data.  For spectro-photometric calibration, we adopted a Kurucz stellar model \citep{Kurucz1979} appropriate for an F3V star, since this spectral type matched HIP 53005's near-IR colors.  Basic processing for NIRC2 data consisted of standard steps for broadband imaging data, including flatfielding, sky subtraction, image registration from directly measuring the (unsaturated) star's position, and photometric calibration.

Visual inspection of the sequence-combined CHARIS and NIRC2 data revealed a point source $\approx$ 0\farcs{}85 from the star in all data sets even without using any PSF subtraction techniques.  Thus, for PSF subtraction we adopted a ``conservative" approach that did not induce severe self-subtraction of the companion's signal.
We mainly utilized ADI to post-process the CHARIS and NIRC2 data sets\footnote{The one exception to this approach is the CHARIS data taken on 2023 February 5th, when we did not obtain sufficient field rotation for ADI. We instead utilized spectral differential imaging \citep[SDI;][]{SDI} to subtract stellar halo and speckles for this data set} (see Table~\ref{tab: observations}).
We tested post-processing using A-LOCI \citep[see][for details]{Currie2018} and {\tt pyKLIP} \citep{pyklip} and the derived photometry and astrometry showed good agreement (see Section~\ref{sec: Results}).
With CHARIS DPP, we produced the reference PSFs at every 10-pix annular area and adopted a moderate ADI post-processing with local pixel masking \citep[see][for details]{Currie2018} to avoid heavy self-subtraction while preserving a high signal-to-noise ratio (SNR). With pyKLIP, we adopted low Karhunen-Lo\`eve (KL) modes \citep[${\rm KL}\leq5$, see][]{Soummer2012} that did not apply aggressive PSF reductions. We did not apply spectral differential imaging to further suppress speckle residuals after ADI.  
We corrected for the (small) signal loss due to PSF subtraction by injecting fake sources, and applied them to correct the extracted spectroscopy (CHARIS) and photometry (NIRC2).

Due to frequent daytime craning events of the SCExAO instruments, it is possible that the detector angle offset of CHARIS for the data taken in early 2022 might have varied from the typical value \cite[$-2\fdg03\pm0\fdg27$;][]{Chen2023,Currie2026}.   
Measuring the north angle offset using different astrometric binaries taken around these epochs resulted in a $1\fdg04$ variation at most from the expected angles. We conservatively included this value in the measurement error of HIP~53005C's position angle specifically for the 2022 February data.
For NIRC2, part of the data were taken with pyWFS and the north angle offset differs from the Shack-Hartmann WFS by $0\fdg118$ \citep{Walker2026}. We corrected this offset in the position angle measurements of HIP~53005C.

\subsubsection{Data Reduction: IRD} \label{sec: Data Reduction: IRD}

\begin{figure}
    \centering
    \includegraphics[width=0.45\textwidth]{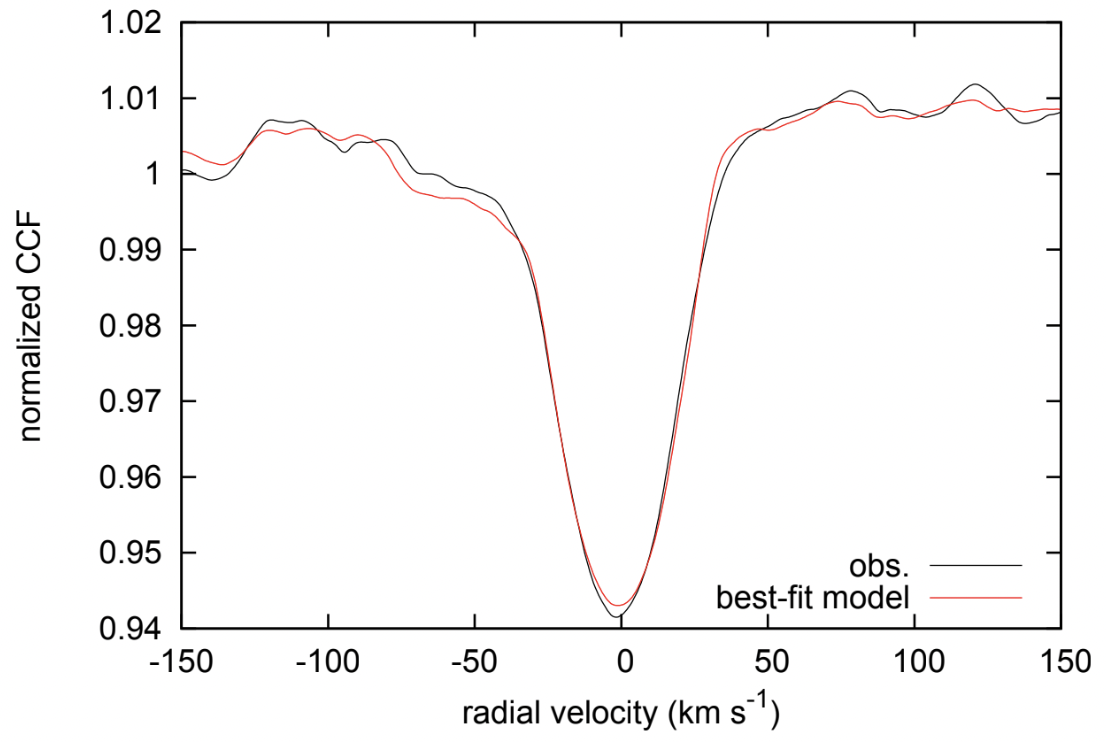}
    \caption{Estimation of the stellar rotation velocity from the IRD spectra.}
    \label{fig: CCF vsini}
\end{figure}

The IRD data were reduced using the IRD pipeline which uses IRAF\footnote{The Image Reduction and Analysis Facility ({\sc IRAF}) is distributed by the US National Optical Astronomy Observatories, operated by the Association of Universities for Research in Astronomy, Inc., under a cooperative agreement with the National Science Foundation.} for the scattered light subtraction, flat fielding and extraction of one-dimensional spectra \citep[][]{Kuzuhara2018,Hirano2020}.  
From the IRD spectra, we estimated the projected rotation velocity $v\sin i$ following the procedure in \citet{Hirano2020}.  Briefly, we first removed telluric lines from the IRD spectra using a rapid rotator's spectrum taken on the same night, and computed the cross-correlating function (CCF) between the observed spectrum and an F0-type spectral template. 
In parallel, we generated simulational data sets by varying $v\sin{i}$ in the F0-type template and incorporating the IRD resolution and a macroturbulence value, which is fixed at $\xi=5.5\ {\rm km~s^{-1}}$ \citep[empirical value at an early-F type from SPOCS;][]{Valenti2005}, and also calculated a CCF of each generated data set. 
We finally implemented a Markov Chain Monte Carlo (MCMC) fitting to explore the best $v\sin{i}$ value in the simulated data sets. The MCMC fitting yields $v\sin{i}=27.11\pm0.59\ {\rm km~s^{-1}}$ (see Figure~\ref{fig: CCF vsini}). Considering possible uncertainties of the stellar template and macroturbulence value, we conservatively adopted $v\sin{i}=27\pm1\ {\rm km~s^{-1}}$.

\begin{figure}
\centering
    \includegraphics[width=0.45\textwidth]{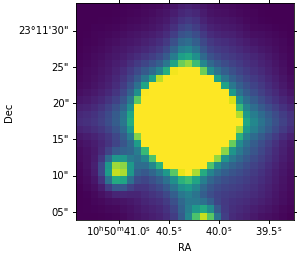}
    \caption{2MASS $J$-band image of the HIP~53005 system. The outer companion (HIP~53005B) is resolved at the southeast. The point-like source located at the south edge is a filter glint artifact. }
\label{fig: HIP53005B}
\end{figure}

\subsection{Archival Data of 2MASS, Pan-STARRS, and Gaia} \label{sec: Archival Data of Space Telescopes}

To identify any very wide-separation companions to HIP 53005, we examined archival 2MASS, Pan-STARRS, and Gaia data.  These data resolved a point source located $\sim12\arcsec$ from HIP~53005 (Figure~\ref{fig: HIP53005B}). 
Gaia astrometry indicates that this point source has a parallax and proper motion consistent with HIP~53005 within the uncertainties, indicating a gravitationally-bound companion; we therefore we denote it HIP~53005B.
Note that the central star in the Pan-STARRS image is heavily saturated and information of the HIP~53005 system includes uncertain systematics, thus we do not include the Pan-STARRS data in further discussions.

\section{System Age} \label{sec: System Age}

\begin{figure}
\centering
    \includegraphics[width=0.45\textwidth]{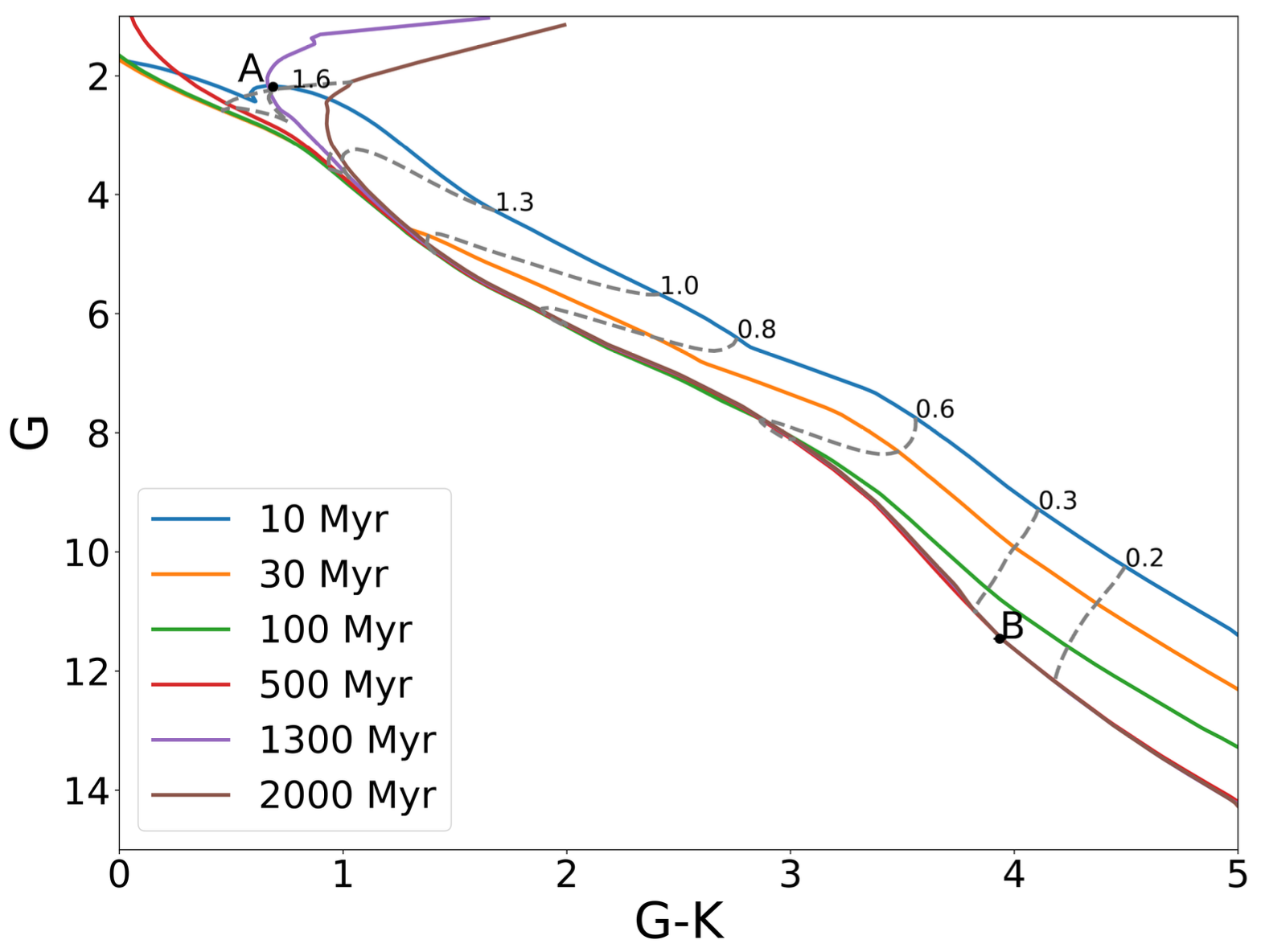}
    \caption{CMD for HIP 53005 A and B. Overplotted, PARSEC tracks and isochrones of different ages and masses. The position of B in the diagram is crucial to break the degeneracy for the system age. }
    \label{fig: CMD}
\end{figure}

To investigate the HIP 53005 system age, we used classical isochrone fitting.
We employed \textsc{madys} \citep{squicciarini22}, a tool that derives stellar or substellar astrophysical parameters (such as mass, age, radius or $T_{\rm eff}$) by comparing appropriate photometric measurements and (sub)stellar evolutionary models.  To achieve these constraints, we dereddened the optical and NIR photometry by integrating the 3D reddening map by \citet{leike20} along the line of sight and then compared these data to predictions from the most recent version of the PARSEC isochrones \citep{nguyen22}.

A degeneracy in absolute magnitude exists between two competing possibilities: 1) a young scenario ($t=10 \pm 1$~Myr); 2) an old scenario ($t=1.3 \pm 0.2$~Gyr), which is consistent with the VOSA fitting using the Geneva2011 isochrone \citep[$\sim$1.15~Gyr;][]{Ekstrom2012}. We independently confirm these results using other evolutionary models such as MIST \citep{dotter16,choi16} and Dartmouth \citep{feiden16}. Accounting for both random uncertainties and systematic differences across models, we derive a mass $M_* = 1.63 \pm 0.03 M_\odot$ and a radius $2.04 \pm 0.06 R_\odot$, with no significant variation between the two scenarios. Although the older solution is favored given that HIP~53005 is not a member of any young association or moving group (see Section~\ref{sec: System Properties}), we looked for additional pieces of evidence to distinguish between the two possibilities.

Starting from the projected rotational velocity estimated in Section~\ref{sec: Data Reduction: IRD} and the stellar radius derived above, we estimated a (maximum) rotation period of $\sim 3.7$~days. Unfortunately, gyrochronology cannot be directly employed for stars earlier than mid-F like HIP~53005 lacking a convective envelope \citep[see, e.g.,][]{gallet19}. As a result, the derived rotational velocity does not support or rule out either of the two scenarios \citep[e.g., Figure~6 in][]{kounkel22}.

The existence of HIP~53005B was found to be crucial to solving the degeneracy. Given the long cooling timescale of M stars, the isochrones corresponding to very young and very old ages are well separated in a color-magnitude diagram (CMD). The photometry of HIP~53005B can only be reconciled with the old scenario (Figure~\ref{fig: CMD}): we therefore adopt for the system an age $t=1.3 \pm 0.2$~Gyr.

\section{Results and Analysis} \label{sec: Results}

\begin{figure*}
\begin{tabular}{cc}
\begin{minipage}{0.5\hsize}
    \centering
    \includegraphics[width=0.95\textwidth]{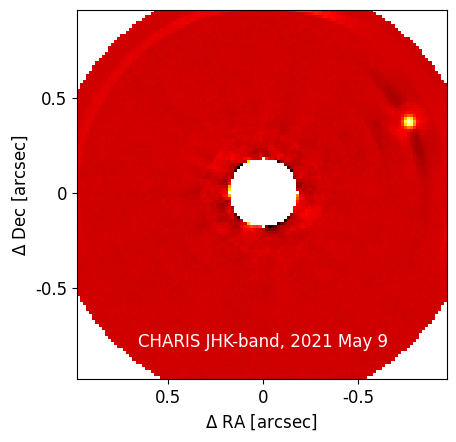}
\end{minipage}
\begin{minipage}{0.5\hsize}
    \centering
    \includegraphics[width=0.95\textwidth]{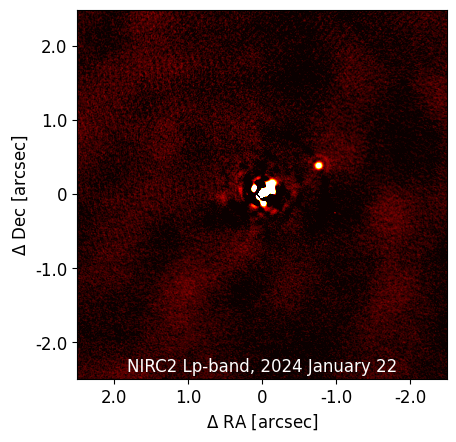}
\end{minipage}
\end{tabular}
    \caption{ADI-reduced CHARIS (left) and NIRC2 (right) images of HIP~53005C. The central star is masked by the algorithm, north is up and east is left.}
    \label{fig: HIP53005C}
\end{figure*}

Our CHARIS and NIRC2 imaging data detected the new companion HIP~53005C (see Figure~\ref{fig: HIP53005C}), while archival Gaia, 2MASS (see Figure~\ref{fig: HIP53005B}), and Pan-STARRS data reveal an outer companion, HIP~53005B. We therefore conclude that HIP~53005 is a multiple system (see Table~\ref{tab: astrometry} in this section).

\subsection{Photometry, Color, and SED}
\label{sec: Photometry, Color, and SED}

\begin{table}
    \centering
    \caption{Photometry and spectrum of HIP~53005C}
\begin{tabular}{c|cccc}
       filter & $J$ & $H$ & $K$ & $L'$  \\ \hline\hline
     contrast & $1.0\times10^{-4}$ & $1.9\times10^{-4}$ & $3.3\times10^{-4}$ & $6.9\times10^{-4}$ \\
     mag & 15.99$\pm$0.03 & 15.23 $\pm$0.03 & 14.51$\pm$0.03 & 13.70$\pm$0.02
\end{tabular}
\begin{tabular}{ccc}
$\lambda$ [$\mu$m] & \multicolumn{2}{c}{Flux [mJy]} \\\hline
&  1st epoch  & 2nd epoch \\ \hline \hline
1.160 & $0.538 \pm 0.021$ & $0.393 \pm 0.057$ \\
1.200 & $0.534 \pm 0.029$ & $0.570 \pm 0.047$ \\
1.241 & $0.587 \pm 0.016$ & $0.602 \pm 0.041$ \\
1.284 & $0.783 \pm 0.019$ & $0.751 \pm 0.046$ \\
1.329 & $0.719 \pm 0.023$ & $0.778 \pm 0.046$ \\
1.375$^*$ & $0.116 \pm 0.025$ & $0.495 \pm 0.034$ \\
1.422 & $0.540 \pm 0.019$ & $0.434 \pm 0.037$ \\
1.471 & $0.550 \pm 0.014$ & $0.520 \pm 0.037$ \\
1.522 & $0.712 \pm 0.016$ & $0.630 \pm 0.036$ \\
1.575 & $0.809 \pm 0.018$ & $0.865 \pm 0.045$ \\
1.630 & $0.944 \pm 0.022$ & $0.926 \pm 0.050$ \\
1.686 & $1.014 \pm 0.021$ & $1.059 \pm 0.051$ \\
1.744 & $0.905 \pm 0.021$ & $0.908 \pm 0.050$ \\
1.805 & $0.750 \pm 0.046$ & $0.839 \pm 0.082$ \\
1.867$^*$ & $0.415 \pm 0.040$ & $0.681 \pm 0.044$ \\
1.932 & $0.788 \pm 0.030$ & $0.722 \pm 0.047$ \\
1.999 & $0.683 \pm 0.018$ & $0.674 \pm 0.043$ \\
2.068 & $0.920 \pm 0.037$ & $0.804 \pm 0.046$ \\
2.139 & $1.083 \pm 0.029$ & $1.022 \pm 0.064$ \\
2.213 & $1.142 \pm 0.037$ & $1.073 \pm 0.081$ \\
2.290 & $0.993 \pm 0.047$ & $0.859 \pm 0.065$ \\
2.369 & $0.908 \pm 0.206$ & $0.868 \pm 0.221$ \\
\end{tabular}
\tablecomments{$^*$ The extracted spectra at these wavelengths are affected by telluric absorption.
}
\label{tab: HIP53005 phot and spectrum}
\end{table}

To further characterize HIP 53005C, we extracted the CHARIS $JHK$-band spectrum and the NIRC2 $L'$-band photometry from the highest-SNR data sets. Table~\ref{tab: HIP53005 phot and spectrum} presents the photometry at these NIR bands and the CHARIS spectra at two epochs in 2021 May and 2022 February, respectively. Note that we do not show the spectrum for the 2023 February data because the spectro-photometric calibration using the latest public version of CHARIS DPP\footnote{See also the github link \url{https://github.com/thaynecurrie/charis-dpp}.} leaves systematics due to the Subaru real-time control system upgrade that took place in late 2022 and changed the header information (private communications).

\begin{figure*}
\begin{tabular}{ccc}
\begin{minipage}{0.333\hsize}
\centering
    \includegraphics[width=0.95\textwidth]{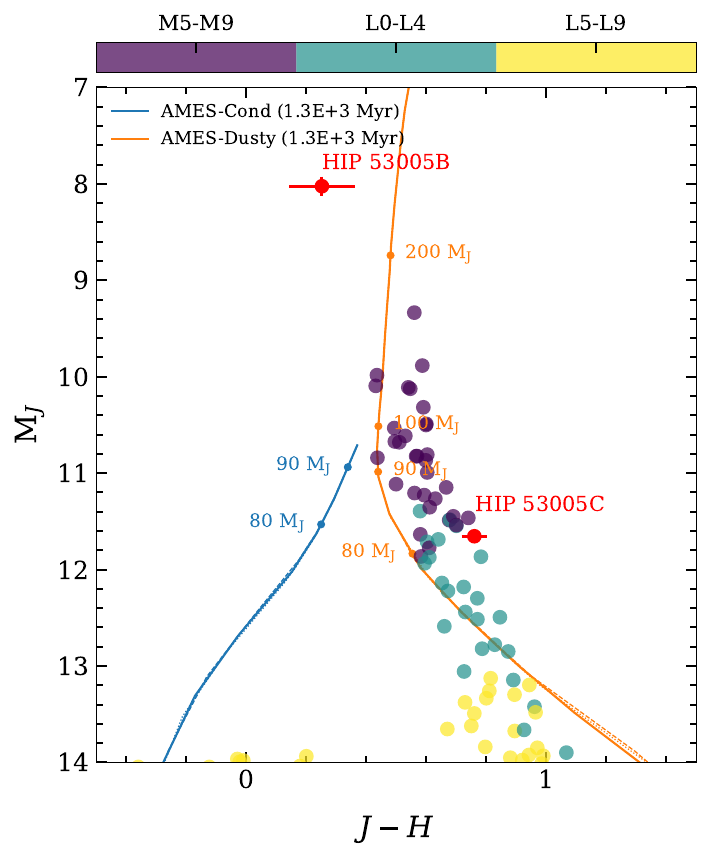}
\end{minipage}
\begin{minipage}{0.333\hsize}
\centering
    \includegraphics[width=0.95\textwidth]{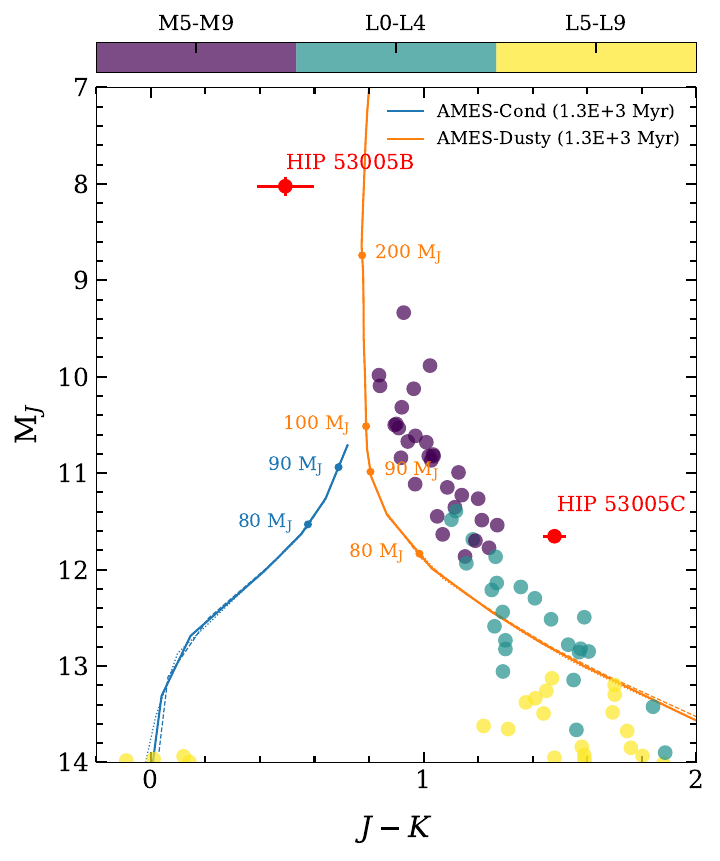}
\end{minipage}
\begin{minipage}{0.333\hsize}
\centering
    \includegraphics[width=0.95\textwidth]{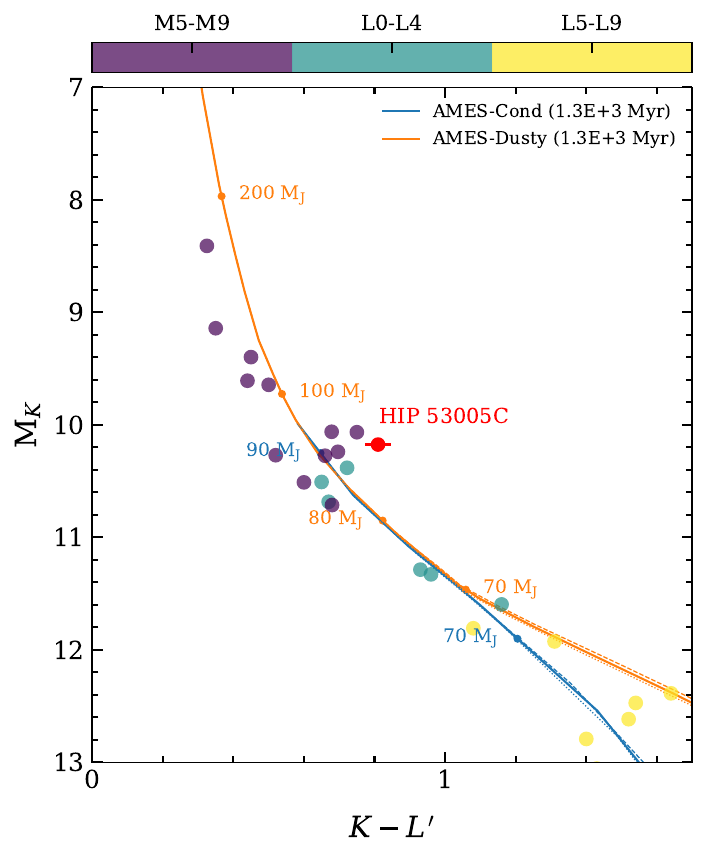}
\end{minipage}
\end{tabular}
\caption{Color-magnitude diagram of the HIP~53005 companions overlaid with the spectral library of low-mass objects (dots) and the AMES isochrones \citep[blue: AMES-Cond, orange: AMES-Dusty;][]{Baraffe2003,Chabrier2000}, made from {\it species}. The solid line and the indicated mass correspond to the best-fit age (1.3~Gyr) and the dashed and dotted lines indicate the age range (dotted line: 1.1~Gyr, dashed line: 1.5~Gyr). Note that the 2MASS catalog does not report the uncertainty of the $J$-band photometry of HIP~53005B, we expediently adopt 0.1~mag for the uncertainty of HIP~53005B's $J$-band magnitude.} We do not have the $L'$-band flux of HIP~53005B and show only C in the $K-L'$ color-magnitude diagram (right).
\label{fig: C-M diagram}
\end{figure*}

\begin{figure*}
\begin{tabular}{cc}
\begin{minipage}{0.55\hsize}
\centering
    \includegraphics[width=\textwidth]{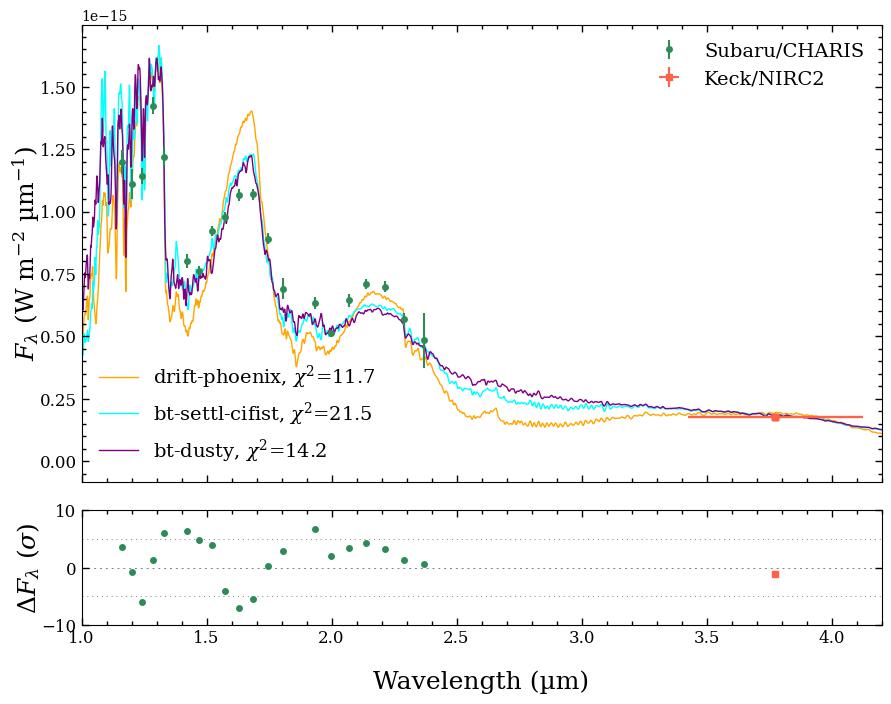}
\end{minipage}
\begin{minipage}{0.45\hsize}
\centering
    \includegraphics[width=\textwidth]{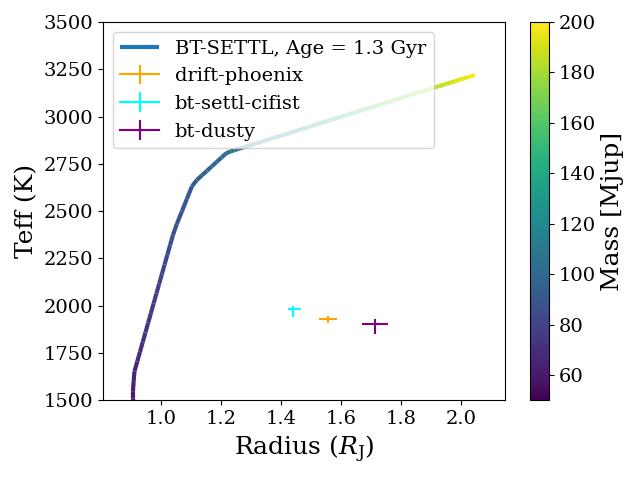}
\end{minipage}
\end{tabular}
\caption{(Left) Best-fit modeled atmosphere within DRIFT-PHOENIX (light orange), BT-SETTL (light blue), and BT-DUSTY (purple) to the observed SED of HIP~53005C with interpolation between the grid points of the models, made with {\tt species}. The spectro-photometric data are overplotted (green for Subaru/CHARIS, orange for Keck/NIRC2), with the low-S/N channels of the CHARIS spectrum excluded from the fit (Table~\ref{tab: HIP53005 phot and spectrum}). The lower panel shows the observed-minus-calculated diagram for the DRIFT-PHOENIX model fitting. 
(Right) Comparison of the derived parameters (radius vs effective temperature) from the best-fit parameters of each model with the BT-SETTL evolutionary model at 1.3~Gyr. The color scale of the evolutionary model indicates mass. 
}
\label{fig: SED fitting}
\end{figure*}

Figure~\ref{fig: C-M diagram} shows the $JHKL'$-band color-magnitude diagram of HIP~53005BC (red) overlaid with a low-mass object library \citep[database of ultracool parallaxes;][]{2012Dupuy,Dupuy2013,Liu2016} and evolutionary models \citep[AMES-Cond (COND03) and AMES-Dusty]{Baraffe2003,Chabrier2000}. HIP~53005B is likely earlier than a mid-M star and HIP~53005C is located at the boundary between the late-M and early-L types. 
As HIP~53005B is a clearly stellar companion \citep[mass$\sim0.22 M_\odot$ based on the $K_{\rm S}$-band flux;][]{Mann2019}, we do not further characterize this companion in detail in this paper.

We further constrained the physical properties of HIP 53005C using two complementary approaches. First, we fitted its SED with atmospheric models (Section~\ref{sec: Deriving physical parameters from atmospheric modeling}). Second, because atmospheric fits near the M/L transition may be systematically biased, we used an empirical dynamical mass--magnitude relation to estimate its mass from the photometry (Section~\ref{sec: Comparison with mass-magnitude diagram}).

\subsubsection{Deriving physical parameters from atmospheric modeling} \label{sec: Deriving physical parameters from atmospheric modeling}

Figure \ref{fig: SED fitting} compares the SED of HIP~53005C with atmospheric models and the best-fit models and parameters. We used the DRIFT-PHOENIX \citep{Witte2011} and BT-SETTL/BT-DUSTY \citep{Allard2012} models, which utilize the {\tt PHOENIX} atmosphere code \citep{Hauschildt1992} but treat dust sedimentation and cloud formation differently to model low-temperature and low-surface gravity objects. All these models have the best $\chi^2$ with $T_{\rm eff}\lesssim2000$.  Assuming a system age of 1.3$\pm0.2$~Gyr and adopting the BT-SETTL evolutionary model, this temperature limit corresponds to mass limit of $<100 M_{\rm Jup}$ (see the right panel in Figure \ref{fig: SED fitting}), but the derived radii seem to be overestimated in all three models.

\subsubsection{Comparison with mass-magnitude diagram}
\label{sec: Comparison with mass-magnitude diagram}
 
\begin{figure}
    \centering
    \includegraphics[width=\linewidth]{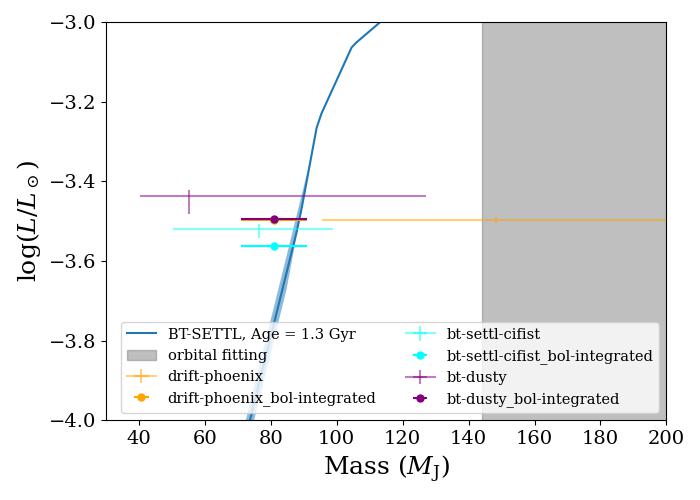}
    \caption{Comparison of the mass \citep[][]{Dupuy2017} and luminosity (integrated bolometric luminosity from the best-fit atmospheric models, with the '\_bol-integrated' suffix in the model label) with the BT-SETTL evolutionary model. The lighter plots indicate the derived parameters from the SED fitting (Section~\ref{sec: Deriving physical parameters from atmospheric modeling}, with the model label). The gray-shaded area indicates a mass range from orbital fitting (Section~\ref{sec: Astrometry and Orbital Fitting}).}
    \label{fig: mass-luminosity}
\end{figure}

We independently estimated the properties of HIP 53005C using the empirical dynamical mass--magnitude relation for ultracool dwarfs \citep[M7--T5;][]{Dupuy2017}. The absolute J- and K-band magnitudes of HIP 53005C (Table~\ref{tab: HIP53005 phot and spectrum}) are consistent with an approximate spectral type of M9--L0 and imply a mass of $81\pm 10~M_{\rm Jup}$. We then combined this empirically inferred mass range with the bolometric luminosity obtained by integrating the best-fit atmospheric model spectrum in Figure~\ref{fig: SED fitting}, and compared the resulting mass--luminosity constraints with the BT-SETTL evolutionary track (Figure~\ref{fig: mass-luminosity}).

Overall, our characterization using HIP~53005C's SED is  consistent with an object at the hydrogen burning limit separating stars from brown dwarfs. However, orbital fitting combining direct imaging and Hipparcos-Gaia proper motion acceleration suggests a higher-mass object (gray area in Figure~\ref{fig: mass-luminosity}, see also Section~\ref{sec: Astrometry and Orbital Fitting}). 
The discrepancy with the orbital fit, together with the radii inferred from the atmospheric models, raises the possibility that HIP 53005 C is an unresolved multiple object.
We discuss potential scenarios to explain the discrepancy with the spectrophotometry-based mass and dynamical mass in Section~\ref{sec: Discussions}.

\subsection{Astrometry and Orbital Fitting} \label{sec: Astrometry and Orbital Fitting}

\begin{deluxetable*}{cccc}
\tablecaption{Summary of Astrometry of the companions around HIP~53005}
\label{tab: astrometry}
\tablewidth{0pt}
\tablehead{
\colhead{Date[UT]} & \colhead{Date [MJD]} & \colhead{Separation [mas]} & \colhead{Position Angle [deg]}
}
\startdata
        \multicolumn{4}{c}{HIP~53005B} \\ 
        \dots & 50842.4491$^a$ & 12419.3$\pm$109.7 & 124.56$\pm$0.48 \\
        \dots & 57023.25$^b$ & 12381.53$\pm$0.46 & 123.759$\pm$0.002 \\
        \dots & 57205.875$^b$ & 12378.83$\pm$0.07 & 123.7485$\pm$0.0003 \\
        \dots & 57388.5$^b$ & 12379.04$\pm$0.04 & 123.7429$\pm$0.0002 \\ \hline
        \multicolumn{4}{c}{HIP~53005C} \\ 
        2021 May 9 & 59343.3 & 842.8$\pm$6.0 & 295.80$\pm$1.20 \\ 
        2021 May 21 & 59355.2 & 837.3$\pm$3.1 & 295.04$\pm$0.21 \\ 
        2022 February 21 & 59630.5 & 847.1$\pm$6.0 & 296.48$\pm$1.57 \\ 
        2022 April 20$^c$ & 59689.3 & 857.4$\pm$4.6 & 294.94$\pm$0.30 \\ 
        2023 February 5& 59980.2 & 845.1$\pm$5.2 & 296.45$\pm$0.28 \\ 
        2024 January 22 & 60331.5 & 851.1$\pm$5.1 & 296.26$\pm$0.34 \\ 
        2025 May 11 & 60806.2 & 855.8$\pm$3.1 & 297.27$\pm$0.20 
\enddata
\tablecomments{
$^a$ 2MASS. $^b$ Based on each epoch of the Gaia data release - DR1, DR2, and DR3. $^c$ Due to potential systematics with the vortex mask we did not include the derived astrometry of this epoch to the orbital fitting. See text for details.}
\end{deluxetable*}

Table~\ref{tab: astrometry} summarizes astrometry of HIP~53005B and C. The B companion is outside the CHARIS and NIRC2 FoVs and we compiled only the archival data of 2MASS and Gaia for B's relative astrometry.
To derive robust astrometry for HIP 53005C, we used the forward modeling technique \citep{FM_Pueyo_2016,Wang2016} implemented in {\tt pyKLIP} to generate a forward model of the companion PSF that accurately captures the distortions created by the Karhunen-Lo\`eve Image Projection (KLIP) algorithm \citep{Soummer2012, Pueyo2015} due to self-subtraction. We then fit this forward model to the PSF subtracted data using MCMC to measure the astrometry of the companion. {\tt pyKLIP} uses Gaussian process regression to account for correlated noise in the imaging data and avoid underestimating the errors. 
For the CHARIS dataset taken on 2023 Feb 5th, there is very little parallactic angle rotation over the entire sequence ($\sim 4^\circ$). Therefore, we modeled the stellar PSF in SDI mode instead of ADI for this epoch, and kept other steps the same. 
We also tested the derived astrometry using MCMC fitting with the A-LOCI reduction \citep{Currie2018} and confirmed consistency within the error bars.

\begin{figure}
    \centering
    \includegraphics[width=\linewidth]{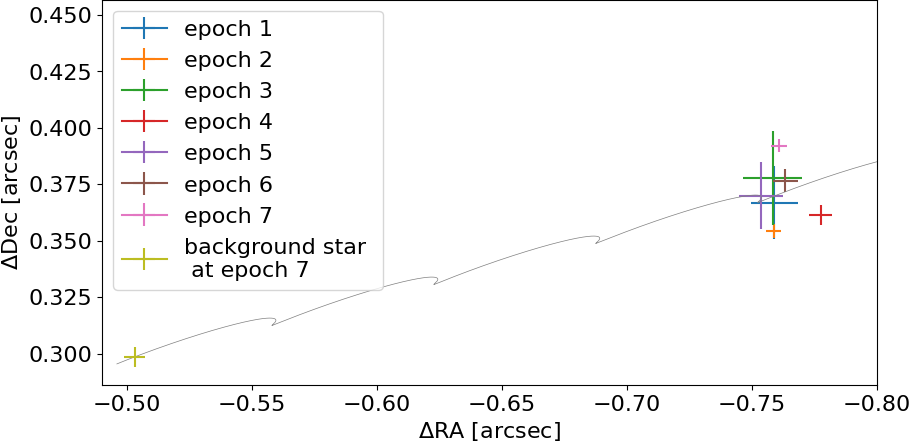}
    \caption{Relative astrometry of HIP~53005C for all the CHARIS and NIRC2 epochs compared with a trajectory assuming a zero proper-motion background star.}
    \label{fig: cpm test}
\end{figure}

Figure~\ref{fig: cpm test} shows the common proper-motion test assuming a background star has a zero proper motion. The latest-epoch relative astrometry is significantly offset from the expected position of a background star case, verifying that HIP~53005C is bound to the central star. The companion's relative astrometry may show some tension with smooth orbital motion, which could be evidence for systematic errors in astrometric calibration or binarity (see Section~\ref{sec: binarity}).

Although HIP~53005B exhibits only limited orbital motion over the available baseline, the measured position angles of both HIP~53005B and HIP~53005C suggest retrograde orbits. Given the wide projected separation of HIP 53005B ($\sim900$~au), a mutual misalignment between the B and C orbits would be consistent with the orbital architectures commonly seen in wide multiple systems \citep{Tokovinin2017}.

\begin{figure*}
\begin{tabular}{ccc}
\begin{minipage}{0.33\hsize}
\centering
    \includegraphics[width=1.05\textwidth]{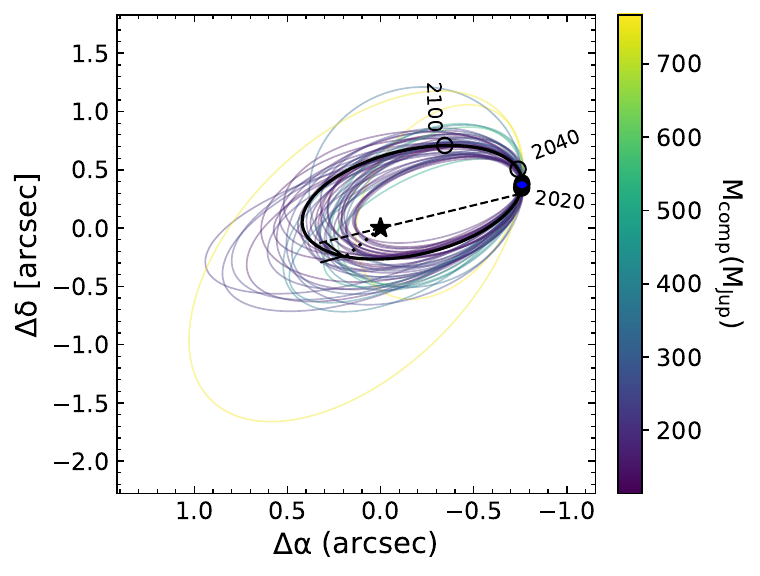}
\end{minipage}
\begin{minipage}{0.33\hsize}
\centering
    \includegraphics[width=\textwidth]{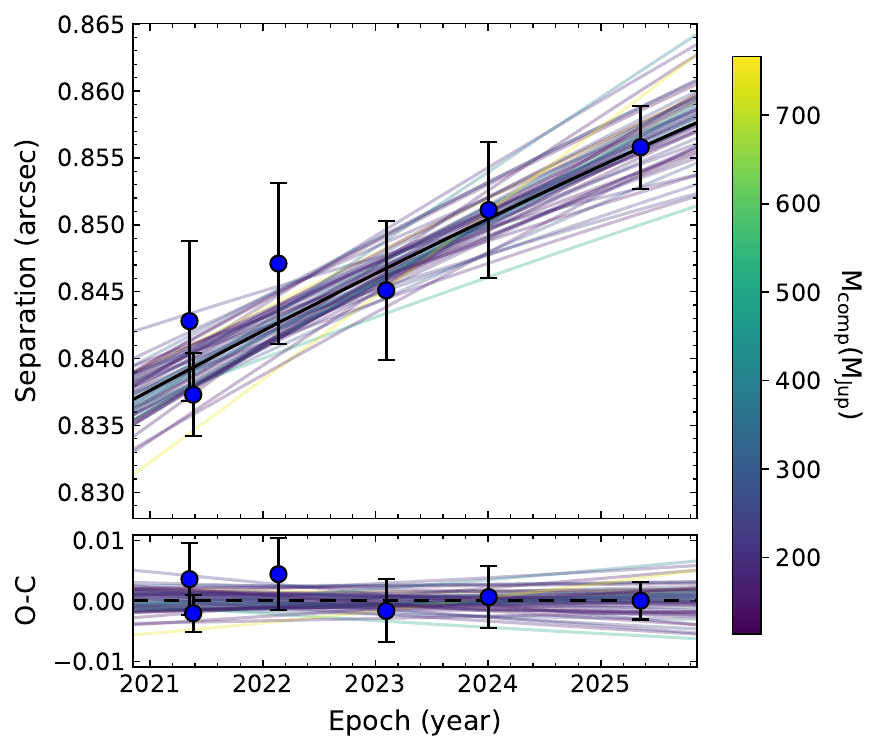}
\end{minipage}
\begin{minipage}{0.33\hsize}
\centering
    \includegraphics[width=\textwidth]{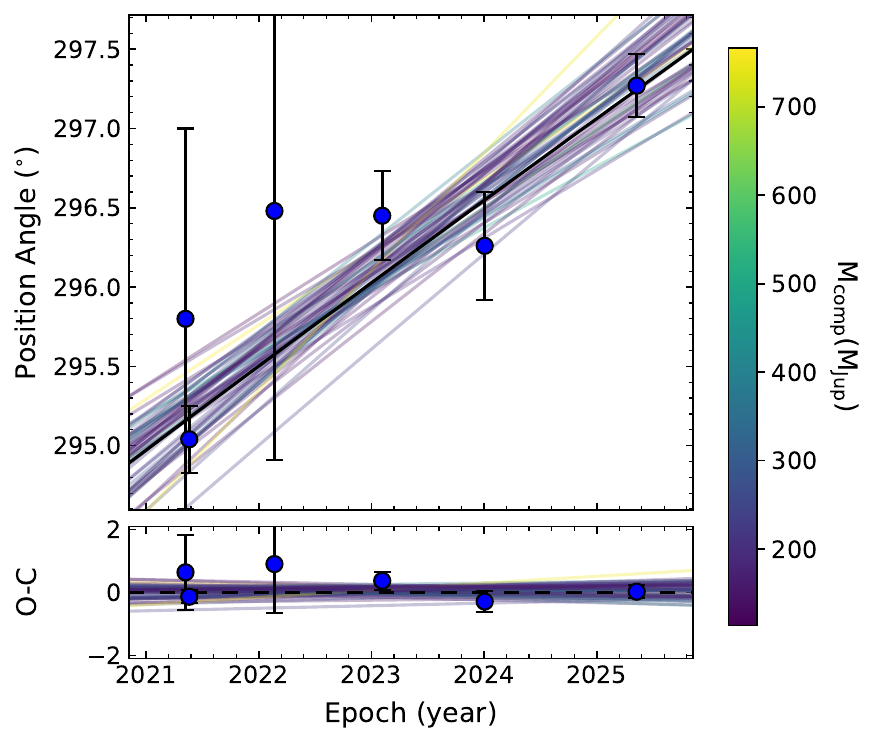}
\end{minipage}
\end{tabular}
\caption{Modeled orbits of the companion HIP~53005C from fitting the Hipparcos-Gaia proper motion acceleration and relative astrometry with Subaru/CHARIS and Keck/NIRC2 (left: projected orbit, middle: separation, right: position angle). The color scale corresponds to the simulated mass of the companion.}
\label{fig: orbit fit results}
\end{figure*}

To estimate HIP~53005C's dynamical mass and other orbital parameters, we used {\tt orvara} \citep{Brandt2021}: a Python-based MCMC code to fit companion relative astrometry and the star's astrometry from the Hipparcos-Gaia Catalog of Accelerations \citep{Brandt2018}. We excluded the 2022 April relative astrometry from the orbital fit because those data were acquired with the vortex coronagraph, which may introduce potential undiagnosed systematics in the image registration processes compared to the other NIRC2 data sets taken without a coronagraph mask.\footnote{Note that we included only HIP~53005C in the orbital fitting.  The B companion's separation is larger by a factor of $\sim$15 ( see Table~\ref{tab: astrometry}).  For a companion mass of $M$ and angular separation of $\rho$, a star's astrometric acceleration scales as $\Delta\mu\propto Ma^{-2}$.  Thus, the B companion's contribution to the star's astrometric acceleration is a factor of $\approx$ 200 lower.  The angle of the proper motion anomaly ($\sim291\fdg5$; $\Delta\mu_{\rm RA}=-0.297\pm0.036\ {\rm mas/yr}$, $\Delta\mu_{\rm Dec}=0.119\pm0.038\ {\rm mas/yr}$) is fully consistent with the position angle of HIP~53005C.}
We adopted default priors for all parameters in {\tt orvara} except the host-star mass ($1.6\pm0.1 M_\odot$).

Table~\ref{tab: orbit fitting} summarizes the orbital fitting results and Figure~\ref{fig: orbit fit results} shows the corner plot displaying the posterior distributions for HIP~53005C's orbital parameters.
Although the fit does not strongly constrain the dynamical mass due to the small orbital motion since the detection, the 1$\sigma$ range (${185}_{-39}^{+116}\ M_{\rm Jup}$) is significantly above the estimated mass from its color-color diagram and the SED fitting result (see Section \ref{sec: Photometry, Color, and SED}).
If HIP~53005C has only the mass implied by its SED, then an additional companion producing a proper-motion acceleration comparable to that induced by HIP~53005C is required to explain the observed signal.

\begin{table}
    \centering
    \caption{Summary of orbital fitting for HIP~53005C}
    \begin{tabular}{cc}
        Parameter & Value  \\ \hline\hline
        Msec ($M_{\rm Jup}$)      &   ${185}_{-39}^{+116}$ \\
        a (AU)     &    ${48.3}_{-7.5}^{+13}$ \\
        inclination (deg)    &   ${45}_{-20}^{+12}$ \\
        ascending node (deg)    &    ${127}_{-25}^{+22}$ \\
        mean longitude (deg)    &    ${125}_{-25}^{+19}$ \\
        period (yrs)     &   
        ${252}_{-56}^{+96}$ \\
        argument of periastron (deg)   &     ${80}_{-58}^{+257}$ \\
        eccentricity    &    ${0.51}_{-0.28}^{+0.18}$ \\
        T0 (JD)    &    ${2515395}_{-11468}^{+22303}$ \\
        mass ratio    &    ${0.111}_{-0.025}^{+0.070}$ \\
    \end{tabular}
    \label{tab: orbit fitting}
\end{table}

\begin{figure*}
    \centering
    \includegraphics[width=0.8\textwidth]{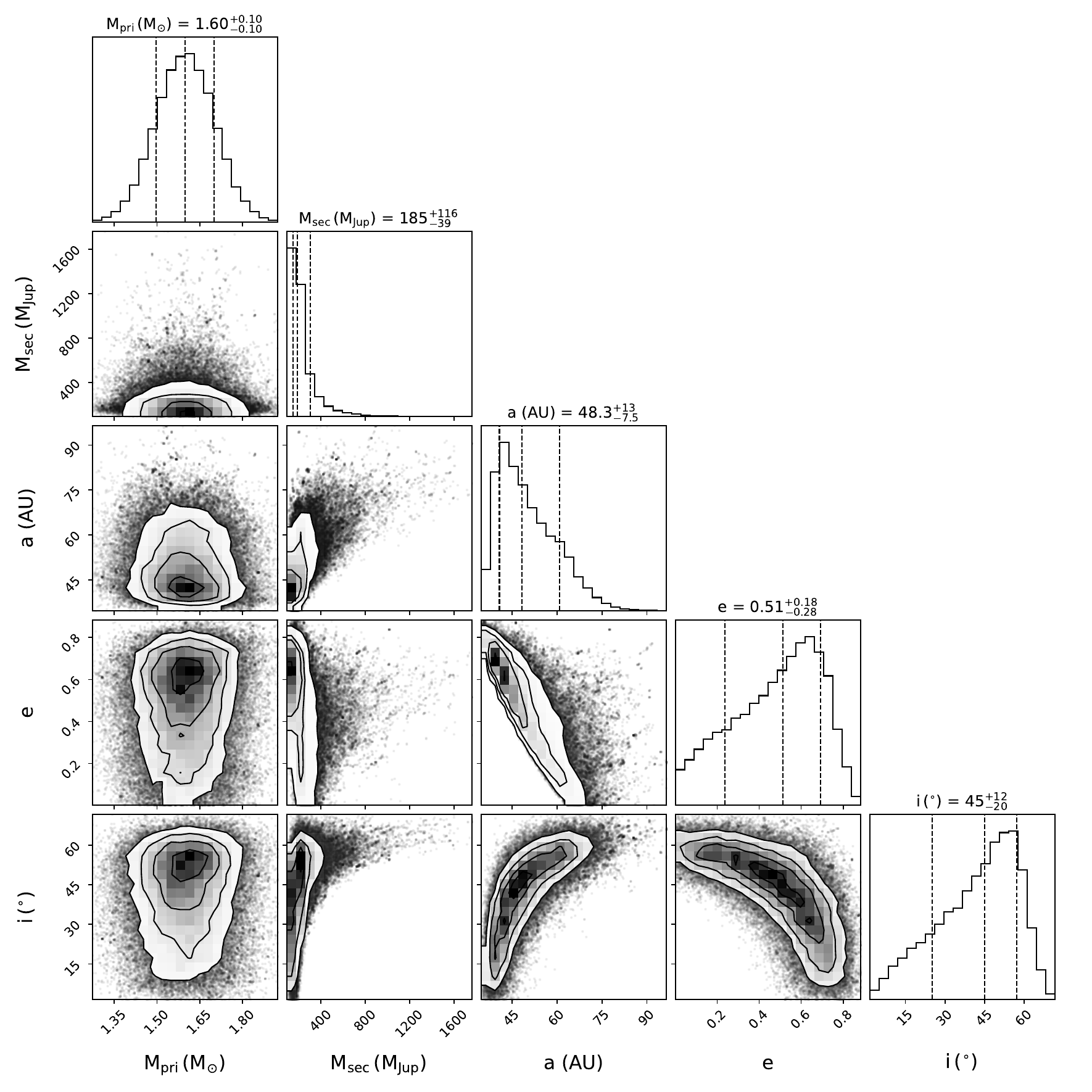}
    \caption{Corner plot of the orbital fitting result.}
    \label{fig: corner plot}
\end{figure*}

\section{Discussion} \label{sec: Discussions}

In Section~\ref{sec: Results}, we obtained different masses for HIP~53005C from its SED and orbital fitting. Other recent studies combining direct imaging with other indirect techniques such as radial velocity or astrometry have reported discrepancies between dynamical masses and those derived from SED modeling for substellar objects \citep[e.g., HD~4113C, HD~47127B, and Gliese~229B;][]{Cheetham2018,Bowler2021,Brandt2020}.
This section mainly discusses potential scenarios to explain this discrepancy. The system age is well constrained by using HIP~53005AB and we do not discuss the uncertainty of the age estimation that could affect characterizations of C's SED.

\subsection{Potential additional companions}

The first potential scenario is unseen companion(s) below the detection limits of our observations (we did not detect additional companion candidates down to $\sim0\farcs1$, see Section \ref{sec: Results}) that can contribute to the Gaia-Hipparcos proper motion acceleration.  
Figure~\ref{fig: contrast limit} shows the CHARIS contrast limit achieving $\sim10^{-5}$ at $0\farcs25$, which is based on the 2021 May data because this data set has the longest total exposure time and achieved the highest contrast to constrain the presence of other objects around the central star. Compared with the $H$-band contrast of HIP~53005C, we could rule out additional sources more massive than C at separations $\geq0\farcs2$.

To further constrain potential companions located close to the CHARIS and NIRC2 inner working angles, we utilized Gaia and TESS data.
Gaia DR3 records RUWE as 1.009 indicating that HIP~53005A is consistent with a single object. Furthermore, there is no record of HIP~53005 in the Gaia DR3 non-single-star acceleration.
We also investigated the TESS light curve and HIP~53005 has been observed in Sectors 22 and 48. We downloaded the raw data and produced lightkurve objects \citep[][data DOI: \url{10.17909/4r8e-f876}]{lightkurve} with the default aperture size, which was processed by removing instrumental effects and outliers that arose from the star crossing at the edge of the TESS FoV.
We did not find any signature of variability larger than 1\% in the TESS lightcurves. 
Also, our IRD spectrum of HIP~53005 did not show any double-line feature suggesting spectroscopic binary.
Considering these conditions mentioned above, it is unlikely that HIP~53005A is a binary but it is still possible that additional substellar-mass object(s) below the CHARIS detection limit can contribute to the proper motion accelerations.

\begin{figure}
    \centering
    \includegraphics[width=0.45\textwidth]{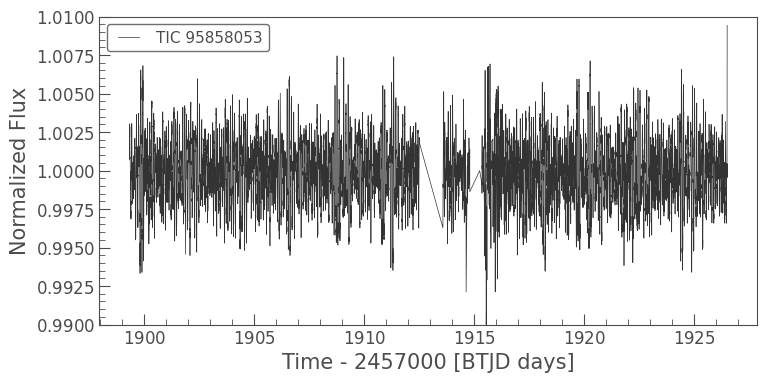}
    \centering
    \includegraphics[width=0.45\textwidth]{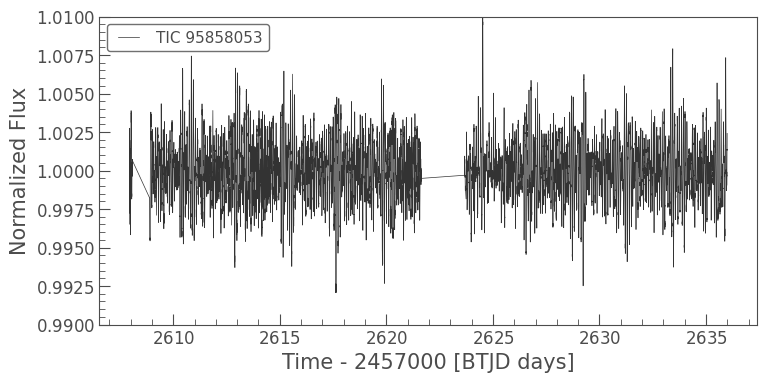}
\caption{TESS light curves of HIP~53005 (TIC 95858053) showing no significant variability on the central star at Sector 22 (top) and 48 (bottom).}
\label{fig: TESS lightcurve}
\end{figure}

\begin{figure}
    \centering
    \includegraphics[width=0.48\textwidth]{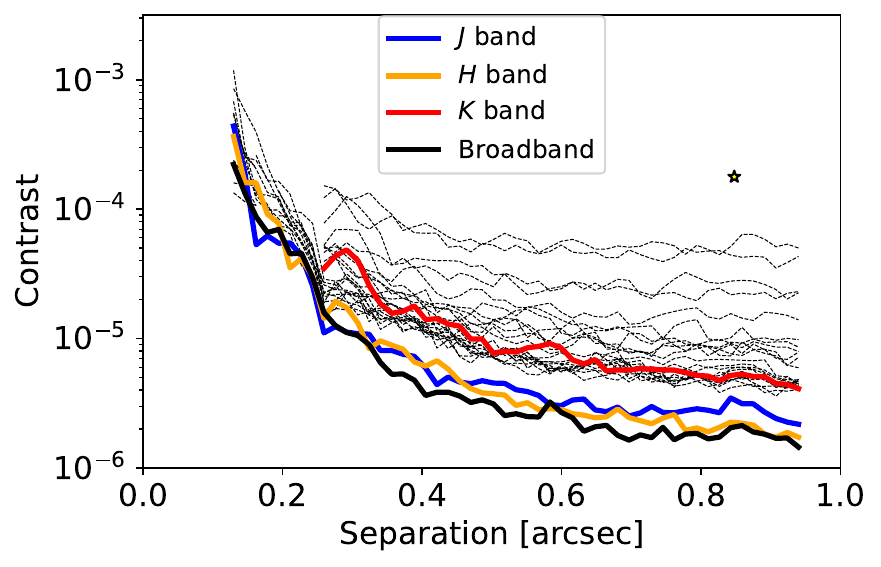}
    \caption{5$\sigma$ contrast limit of the CHARIS ADI+SDI reduction. The dashed lines correspond to the contrast limit at each CHARIS channel and the solid lines correspond to that from the combined image at $J$, $H$, $K$, and the full CHARIS channels. The star symbol indicates HIP~53005C with the $H$-band contrast.}
    \label{fig: contrast limit}
\end{figure}

\subsection{Is HIP~53005C itself a binary?} \label{sec: binarity}

\begin{figure*}
\centering
    \includegraphics[width=0.85\textwidth]{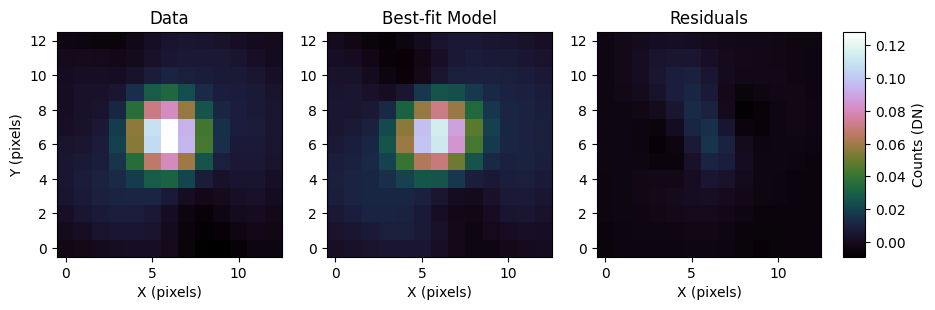}
\caption{Best-fit forward modeled PSF to the CHARIS HIP 53005 C data taken on May 9, 2021. The left panel shows the ADI reduced CHARIS data, the middle panel shows the best-fit forward model and the right panel shows the residuals. The data reduction and astrometry fitting are carried out using pyKLIP \citep{pyklip}.}
\label{fig: CHARIS psf fit}
\end{figure*}

\begin{figure*}
\centering
    \includegraphics[width=0.85\textwidth]{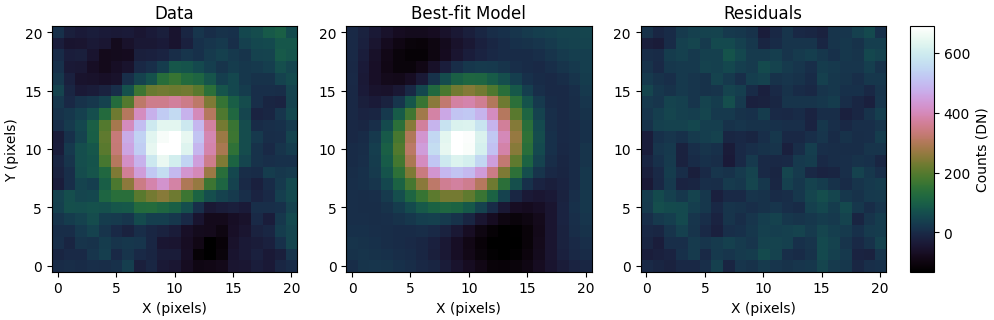}
\caption{Same as Figure~\ref{fig: CHARIS psf fit} except that the data were taken by NIRC2 on January 22, 2024.}
\label{fig: NIRC2 psf fit}
\end{figure*}

Another possibility is that HIP~53005C is itself a multiple system, similar to Gliese 229B \citep[Gliese~229Bab;][]{Brandt2020,Xuan2024,Whitebook2024}, which would naturally explain the discrepancy between the SED-based and dynamical mass estimates. As found in Section~\ref{sec: Deriving physical parameters from atmospheric modeling} our derived radius is substantially larger than evolutionary model predictions, potentially an indicator that HI~P53005C itself is a multiple.
As mentioned in Section~\ref{sec: Astrometry and Orbital Fitting}, fluctuating relative astrometry might be caused because of an unresolved binary, whose PSF centroid can be variable by its 'binary orbit'. In this scenario, the {\tt orvara} fitting does not take into account for the binary orbital motion and the derived parameters from orbital fitting may be different from the 'true' values.
We investigated if we see additional sources after forward modeling, and Figures~\ref{fig: CHARIS psf fit} and \ref{fig: NIRC2 psf fit} show the best-fit forward-modeling results with the CHARIS and NIRC2 data where HIP~53005C was detected at the highest SNR (2021 May data for CHARIS, 2024 Jan data for NIRC2), respectively. 
The PSF of HIP~53005C was well fit with the instrumental PSF and we did not see any significant features in the residual map. The other-epoch data did not show any features suggesting additional sources.
If C is a binary, it should be a very tight system whose separation is smaller than the resolution of our CHARIS and NIRC2 observations ($<0\farcs1$).

\begin{figure*}
    \centering
    \includegraphics[width=0.9\textwidth]{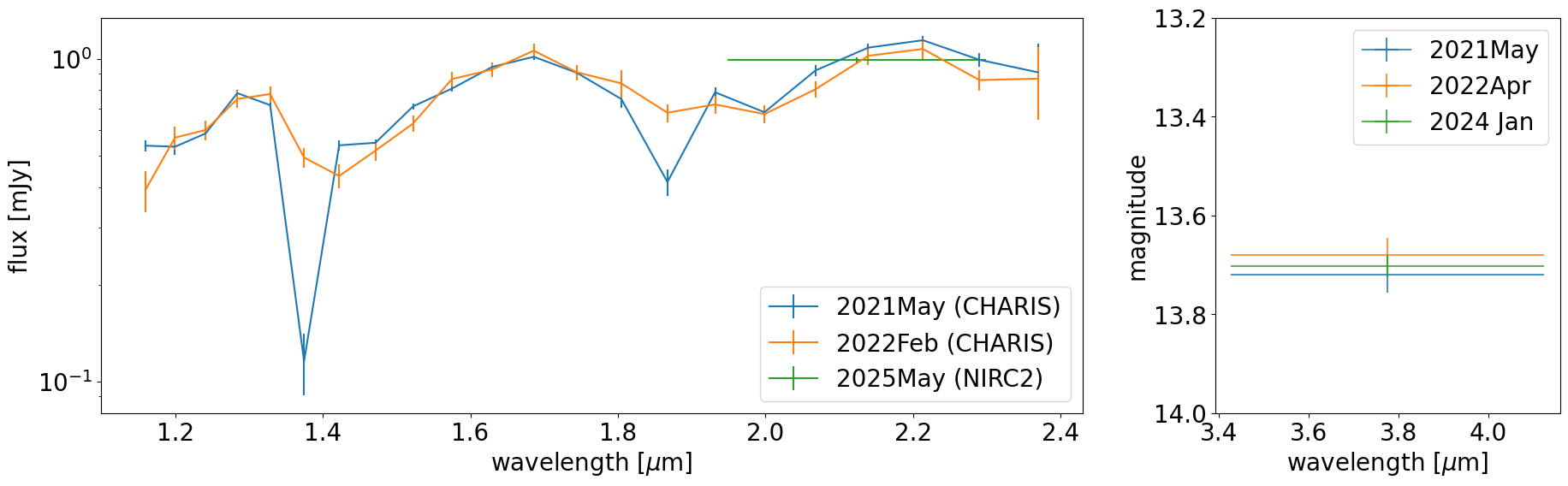}
    \caption{Comparison of the CHARIS $JHK$-band spectra and the NIRC2 $K'$-band photometry (left) and NIRC2 $L'$-band photometry (right) at different epochs. Note that the dips at $\sim1.4\ \micron$ and $\sim1.9\ \micron$ in the 2021 CHARIS spectrum are due to atmospheric absorptions of the Earth. }
    \label{fig: check variability}
\end{figure*}

We also compared the CHARIS spectra taken in 2021 May and 2022 February and the NIRC2 photometry among 2021 May, 2022 April, and 2024 January by incorporating all the CHARIS channels except for those at $1.329-1.422, 1.805-1.932~{\rm \mu m}$, where fluxes are affected by the atmospheric absorptions of the Earth (see Table~\ref{tab: HIP53005 phot and spectrum} and Figure \ref{fig: check variability}), and we did not recognize significant variation in the infrared flux ($\sim2.37\sigma$). As mentioned in Section~\ref{sec: Photometry, Color, and SED}), we did not use the CHARIS data taken in 2023 February to test the variability because of the systematics in the header.

\section{Summary} \label{sec: Summary}

We present Subaru/CHARIS and Keck/NIRC2 high-contrast imaging data detecting a new companion (HIP~53005C) around an early-type star HIP~53005, motivated by a previous detection of a proper motion acceleration from Hipparcos and Gaia astrometry.  Additionally, archival data reveal a very wide-separation stellar companion (HIP~53005B), indicating that this system is a multiple system.
We obtained a CHARIS $JHK$-band spectrum and Keck/NIRC2 $K'L'$-band photometry, and multi-epoch astrometry. 

The color-magnitude diagrams, atmospheric-model fits, and empirical mass-magnitude relation all suggest that HIP 53005 C lies at the M/L spectral-type transition, with an estimated mass of $\sim80\pm10~M_{\rm Jup}$. However, the dynamical mass derived from the orbital fitting that combines direct-imaging relative astrometry with Hipparcos-Gaia proper motion acceleration resulted in much higher mass (${185}_{-39}^{+116}\ M_{\rm Jup}$). We investigated two possible scenarios: 1) an additional companion below the detection limit (less massive than HIP~53005C at $\rho\lesssim0\farcs2$) but also contributing to the system acceleration, or 2) HIP~53005C itself being a low-mass binary like Gliese~229Bab \citep{Xuan2024}, but did not confirm or rule out either of these scenarios. This system is a potentially intriguing system to study a multiple star formation, and future follow-up spectroscopic observations such as Subaru/REACH or Keck/KPIC (or Keck/HISPEC) will enable to identify the nature of this object. 

\section*{Acknowledgements}

The authors would like to thank the anonymous referee for their constructive comments and suggestions to improve the quality of the paper.
T.C. was supported by National Science
Foundation (NSF) Astronomy and Astrophysics grant \#2408647. J.W.X is thankful for support from the Heising-Simons Foundation 51 Pegasi b Fellowship (grant \#2025-5887). M.K. was supported by JSPS KAKENHI Grant Number 24K07108. Y.H. was supported by JSPS KAKENHI Grant Number 24H00017.

This research is based on data collected at the Subaru Telescope, which is operated by the National Astronomical Observatory of Japan. Data analysis was in part carried out on the Multi-wavelength Data Analysis System operated by the Astronomy Data Center (ADC), National Astronomical Observatory of Japan.
The development of SCExAO was supported by JSPS (Grant-in-Aid for Research \#23340051, \#26220704, and \#23103002), Astrobiology Center of NINS, Japan, the Mt Cuba Foundation, and the director’s contingency fund at Subaru Telescope. CHARIS was developed under support from the Grant-in-Aid for Scientific Research on Innovative Areas \#2302. SCExAO’s adaptive optics loops and high-speed data acquisition are handled by the CACAO package, which is supported by NSF award 2410616.
Part of data presented herein were obtained at the W. M. Keck Observatory, which is operated as a scientific partnership among the California Institute of Technology, the University of California and the National Aeronautics and Space Administration. The Observatory was made possible by the generous financial support of the W. M. Keck Foundation. This research has made use of the Keck Observatory Archive (KOA), which is operated by the W. M. Keck Observatory and the NASA Exoplanet Science Institute
(NExScI), under contract with the National Aeronautics and
Space Administration. 
The authors wish to recognize and acknowledge the very significant cultural role and reverence that the summit of Maunakea has always had within the indigenous Hawaiian community.  We are most fortunate to have the opportunity to conduct observations from this mountain.

This work has made use of data from the European Space Agency (ESA) mission
{\it Gaia} (\url{https://www.cosmos.esa.int/gaia}), processed by the {\it Gaia}
Data Processing and Analysis Consortium (DPAC,
\url{https://www.cosmos.esa.int/web/gaia/dpac/consortium}). Funding for the DPAC has been provided by national institutions, in particular the institutions participating in the {\it Gaia} Multilateral Agreement.
Funding for the TESS mission is provided by NASA’s Science
Mission Directorate. We acknowledge the use of public TESS
data from pipelines at the TESS Science Office and at the TESS Science Processing Operations Center (SPOC). Resources
supporting this work were provided by the NASA High-End
Computing (HEC) Program through the NASA Advanced
Supercomputing (NAS) Division at Ames Research Center for
the production of the SPOC data products. This paper includes data collected by the TESS mission that are publicly available from the Mikulski Archive for Space Telescopes (MAST).
This research made use of Lightkurve, a Python package for Kepler and TESS data analysis \citep{lightkurve}.
This research has made use of NASA's Astrophysics Data System Bibliographic Services.
This research has made use of the SIMBAD database, operated at CDS, Strasbourg, France.



\bibliography{library}                                    
\end{document}